\documentclass{sig-alternate}
\usepackage{amsmath}
\usepackage{algorithm}
\usepackage{algpseudocode}

\begin{document}
\title{Twitter Hashtag Recommendation using Matrix Factorization}
%
%
%
%
%

\numberofauthors{2} 
%
\author{
%
%
Hamidreza Alvari\\
       \affaddr{Department of EECS, University of Central Florida, Orlando, FL, USA}\\
       \email{halvari@eecs.ucf.edu}
}

\maketitle
\begin{abstract}
Twitter, one of the biggest and most popular microblogging Websites, has evolved into a powerful communication platform which allows millions of active users to generate huge volume of microposts and queries on a daily basis. To accommodate effective categorization and easy search, users are allowed to make use of hashtags, keywords or phrases prefixed by hash character, to categorize and summarize their posts. However, valid hashtags are not restricted and thus are created in a free and heterogeneous style, increasing difficulty of the task of tweet categorization. In this paper, we propose a low-rank weighted matrix factorization based method to recommend hashtags to the users solely based on their hashtag usage history and independent from their tweets' contents. We confirm using two-sample $t$-test that users are more likely to adopt new hashtags similar to the ones they have previously adopted. In particular, we formulate the problem of hashtag recommendation into an optimization problem and incorporate hashtag correlation weight matrix into it to account for the similarity between different hashtags. We finally leverage widely used matrix factorization from recommender systems to solve the optimization problem by capturing the latent factors of users and hashtags. Empirical experiments demonstrate that our method is capable to properly recommend hashtags. 
\end{abstract}




\section{Introduction}
Twitter\footnote{https://twitter.com} is one of the prevalent and well-known microblogging Websites with millions of active users interacting with each other and posting tweets, a message up to 140 characters, per day on either computers or mobile devices. It is popular for massive spreading of tweets and the nature of freedom. Daily bursts of news, gossips, rumors, discussions and many others are all exchanged and shared by users all over the world, no matter where they come from, civilized or uneducated, or even what religion they hold. Consequently, users on Twitter are easily overwhelmed by the tremendous volume of data. The proliferation of such an unstructured user-generated data as opposed to the traditional structured data, has enabled researchers to study and analyze human behavior and develop complex systems such as hashtag recommendation systems that has recently drawn few researchers attention. 

On Twitter, users are freely allowed to assign valid hashtags to their tweets, i.e. strings prepended with the hash "\#" character, to categorize their posts and represent a coarse-grained topic of the content. Hashtags are indeed a community-driven convention for adding additional context to tweets. This mechanism helps tweet search and quickly propagation of the topic among millions of users by allowing them to join the discussion. Although facilitating the task, the heterogeneity and arbitrariness of the hashtags, due to the fact that users do not face any restrictions while creating them, can immediately make mess and hence make subsequent searches for tweets difficult. 

To tackle the problem, few approaches has been proposed in the literature \cite{conf/www/GodinSNSW13, kywe2012recommending,li2011twitter, conf/www/She014, zangerle2011recommending}, most of them rely on either tweet similarity, i.e. content similarity or focus on adopting Latent Drichlet Allocation (LDA) \cite{Blei:2003:LDA:944919.944937} and Latent Semantic Indexing (LSI)~\cite{Deerwester90indexingby}  to capture the abstracted topics of tweets. The problem with these approaches is that not all the time, the contents of tweets are available due to the fact that many users do not usually have public timelines. Even if all users' timelines were public, these methods suffer from computation overloads and lack of strong natural language processing techniques. As opposed to the existing methods, in this paper, we rather take another simple yet effective approach and propose a new hashtag recommendation method which works only based on the users' interests and their hashtag usage history without incorporating the network structure and using tweet/user similarity or LDA.

In more details, we treat the problem of hashtag recommendation as an optimization problem and solve it via widely employed matrix factorization borrowed from recommendation systems. Moreover, we envision that users are more likely to adopt similar hashtags while posting on Twitter and to verify that, we use two-sample $t$-test. Our contributions are thus summarized as follows:

\begin{itemize}
\item We perform two-sample \textit{t}-test to verify that users keep adopting related and similar hashtags and hence possess consistent hashtag usage history.  

\item We address the hashtag recommendation problem with an optimization problem and propose a weighted matrix factorization based method $hWMF$ to recommend hashtags to the users. To ease the process of optimization, we use alternating least square scheme for updating the corresponding matrices as finding the optimal values for them is tedious.

\item We integrate the concept of correlation between the related hashtags into the matrix factorization equation. We quantify the correlation between hashtags based on the times they have been used together and incorporate these values into the optimization problem to weigh the contribution of training samples over test samples.

\item We collect and build a dataset of tweets which contains at least five hashtags very close to a predefined list of 25 trending hashtags (hereafter called seeds list) selected from different categorizes. We evaluate the model on this dataset and demonstrate its ability to recommend hashtags that best aligns with the users' hashtag usage history.
\end{itemize}

The remainder of the paper is organized as follows. In Section 2, we first define the problem of interest and then detail our proposed matrix factorization method to solve the problem. We explain the experimental settings and discuss the results in Section 3. Then, we discuss the related work in Section 4 and conclude the paper in Section 5.
    
\section{Proposed Framework}
In this section, prior to describing the proposed method, we first provide the formal definition of the problem and the notations used throughout the paper. Then we detail the proposed model $hWMF$ for Twitter hashtag recommendation and explain the time complexity of the method. We finally describe two sample $t$-test to verify our initial assumption about users' hashtag usage.

\subsection{Problem Statement}
Given a set of $N$ users and a set of $M$ hashtags adopted by them, we aim to recommend the most related hashtags to the users based on their interests and the hashtag usage history. Suppose we have a very sparse and low-rank user-hashtag matrix $\textbf{X}=[x_{ij}] \in \mathbb{R}_{+}^{N\times M}$. We denote by $x_{ij}, 1\leq i \leq N, 1\leq j \leq M$ the $i$th row and $j$th column of $\textbf{X}$, which represents if user $i$ has adopted hashtag $j$, if $x_{ij}=1$, but has nothing to say otherwise. The zeros in the matrix demonstrate unknown or missing values; a user might have adopted a hashtag in the past but we could not figure it out since our data collection method does not always return all data. The reason that $\textbf{X}$ is sparse is because in Twitter, users do not usually adopt hashtags while tweeting and despite the availability of this feature, only 8\% of the tweets contain hash "\#" character \cite{kywe2012recommending} Also, suppose we have a hashtag-hashtag matrix $\textbf{Y}=[y_{ij}] \in \mathbb{R}_{+}^{M\times M}$ with $y_{ij},  1\leq i \leq M, 1\leq j \leq M$ equals to the number of times hashtags $i$ and $j$ are used together.

We treat the problem of hashtag recommendation as a collaborative filtering based one and formulate it into an optimization problem and employ low-rank matrix factorization to solve it as this method has been widely and successfully employed in various applications such as collective filtering \cite{koren2008factorization} and document clustering \cite{Zhu07}. In its basic form and in the context of recommendar systems, matrix factorization, one of the realizations of latent factor models, captures both items and users by vectors of factors inferred from the ratings. Here, instead we characterize users and hashtags by inferring vectors of factors from user's hashtag usage history. 

\subsection{Matrix factorization model}
Based on matrix factorization scheme, we seek two low-rank and non-negative matrices $\textbf{U} \in \mathbb{R}_{+}^{N\times d}$ and $\textbf{V} \in \mathbb{R}_{+}^{M\times d}$ with dimensionality of the latent space $d \ll M,N$ via solving the following optimization problem:

\begin{equation} \label{eq1} min_{\textbf{U},\textbf{V}} ||\textbf{W} \odot (\textbf{X} - \textbf{UV}^T)||^2_F + \gamma_1||\textbf{U}||^2_F + \gamma_2||\textbf{V}||^2_F\end{equation}

where $\odot$ is Hadamard product (element-wise product) where $(\textbf{X}\odot \textbf{Y})_{ij} = \textbf{X}_{ij} \times \textbf{Y}_{ij}$ for any two matrices $\textbf{X}$ and $\textbf{Y}$ with the same size, $||\textbf{.}||_F$ is the Frobenius norm of a matrix, $||\textbf{A}||_F=\sqrt{\sum_i \sum_j \textbf{A}_{ij}^2}$ and $\textbf{W}=[w_{ij}] \in \mathbb{R}_{+}^{N\times M}, 1\leq i \leq N, 1\leq j \leq M$ is an indicator matrix (i.e. weight matrix) to control the learning process. Also, $\gamma_1 >0 $ and $\gamma_2 > 0$ are non-negative regularization parameters and $||\textbf{U}||^2_F$ and $||\textbf{V}||^2_F$ are two smoothness regularization terms to avoid overfitting. The row vectors $u_{i.}, 1\leq i \leq N$ and  $v_{j.}, 1\leq j \leq M$ denote the low-dimensional representations of users and hashtags respectively.

We integrate $\textbf{W}$ into the optimization equation to avoid impacts of unknown elements of $\textbf{X}$, i.e. increase the contribution of the elements with known values in the optimization process over the elements with the missing information. Therefore, for those hashtags that we have information for, i.e. whether a user has adopted them before, we use $w_{ij}=1$ and for those with missing information we use the average of their correlation with other hashtags with information. In other words, the indicator matrix is formally defined as: 

\begin{equation}\label{eq:w}
w_{ij}=\begin{cases}
1, & x_{ij} = 1\\
\it{\frac{\sum_{j,k} corr(h_j,h_k)}{|A|}}, & x_{ij} = 0, x_{ik} = 1, y_{jk} \geq 1
\end{cases}
\end{equation}

where $A=\{j\mid \forall_{j}, x_{ij}=1\}$ and $corr(h_j,h_k)$, i.e. correlation between hashtags $h_j$ and $h_k$, is calculated by:

\begin{equation}\label{eq:corr}
corr(h_k,h_j) = \frac{y_{jk}}{\sum_{t \ne k} y_{jt}}
\end{equation} 

\textbf{Optimization.} The coupling between $\textbf{U}$ and $\textbf{V}$ in the optimization problem, makes it difficult to find the optimal solutions for both matrices. Therefore, in this work, we adopt the alternating least squares method\cite{conf/icdm/DingLJ08} to solve the optimization problem, where the objective function is iteratively optimized with respect to one of the variables $\textbf{U}$ and $\textbf{V}$ while fixing the other one until convergence. Optimizing the equation ~\ref{eq1} with respect to $\textbf{U}$ and $\textbf{V}$ corresponds to the computation of their derivatives via the following equations. Given the following objective function:

\begin{equation}
\textbf{L} = ||\textbf{W} \odot (\textbf{X} - \textbf{UV}^T)||^2_F + \gamma_1||\textbf{U}||^2_F + \gamma_2||\textbf{V}||^2_F
\end{equation}

the update equations for $\textbf{U}$ and $\textbf{V}$ are computed according to the following equations:
\begin{equation}\label{eq:UUpdate}
\textbf{U} = \textbf{U} - \lambda \frac{\partial \textbf{L}}{\partial \textbf{U}}
\end{equation}
\begin{equation}\label{eq:VUpdate}
\textbf{V} = \textbf{V} - \lambda \frac{\partial \textbf{L}}{\partial \textbf{V}}
\end{equation}
where $\lambda > 0$ is the learning step and the partial derivatives of $\textbf{L}$ with respect to $\textbf{U}$ and $\textbf{V}$ are then obtained using:

\begin{equation}\label{eq:partialU}
\frac{\partial \textbf{L}}{\partial \textbf{U}} = -2\textbf{(W}\odot \textbf{X)}\textbf{V} + 2\textbf{(W}\odot(\textbf{UV}^T))\textbf{V} + \gamma_1\textbf{U}
\end{equation}
\begin{equation}\label{eq:partialV}
\frac{\partial \textbf{L}}{\partial \textbf{V}} = -2(\textbf{W}\odot \textbf{X})^T\textbf{U} + 2(\textbf{W}\odot(\textbf{UV}^T))^T\textbf{U} + \gamma_2\textbf{V}
\end{equation}

Upon the convergence, we approximate $\textbf{X}$ by multiplying the low-rank matrices $\textbf{U}$ and $\textbf{V}$:
\begin{equation}
\widetilde{\textbf{X}} = \textbf{UV}^T
\end{equation}

\textbf{Algorithm.} The detailed algorithm for the proposed matrix factorization framework is shown in Algorithm~\ref{alg:alg1}. After randomly initializing matrices $\textbf{U}$ and $\textbf{V}$ and constructing the indicator matrix $\textbf{W}$ in lines 3-5, in lines 7-10, we alternatingly update $\textbf{U}$ and $\textbf{V}$ based on the equations~\ref{eq:UUpdate} to~\ref{eq:partialV}, until we reach convergence. Practically, convergence is achieved whenever predefined maximum number of iterations has been reached or there is little change in the objective function value. Finally,  $\widetilde{\textbf{X}} = \textbf{UV}^T$ is the low-rank representation of user-hashtag matrix $\textbf{X}$ and also is non-negative as $\textbf{U}$ and $\textbf{V}$ are both non-negative matrices. 

\begin{algorithm}
\caption{\textbf{The proposed framework hWMF}}\label{alg:alg1}
\begin{algorithmic}[1]
\State \textbf{Input:} User-hashtag matrix $\textbf{X}$, hashtag-hashtag matrix $\textbf{Y}$, $d$, $\gamma_1$, $\gamma_2$, $\lambda$
\State \textbf{Output:} Modeled matrix $\widetilde{\textbf{X}}$
\State \text{Initialize \textbf{U} randomly}
\State \text{Initialize \textbf{V} randomly}
\State \text{Construct the indicator matrix $\textbf{W}$ according to eq.~\ref{eq:w}}
\While{\text{Not convergent}}
\State $\frac{\partial \textbf{L}}{\partial \textbf{U}} = -2\textbf{(W}\odot \textbf{X)}\textbf{V} + 2\textbf{(W}\odot(\textbf{UV}^T))\textbf{V} + \gamma_1\textbf{U}$
\State \text{Update} $\textbf{U}\gets \textbf{U} - \lambda  \frac{\partial \textbf{L}}{\partial \textbf{U}}$
\State $\frac{\partial \textbf{L}}{\partial \textbf{V}} = -2(\textbf{W}\odot \textbf{X})^T\textbf{U} + 2(\textbf{W}\odot(\textbf{UV}^T))^T\textbf{U} + \gamma_2\textbf{V}$
\State \text{Update} $\textbf{V}\gets \textbf{V} - \lambda  \frac{\partial \textbf{L}}{\partial \textbf{V}}$
\EndWhile
\State \text{Set $\widetilde{\textbf{X}} = \textbf{UV}^T$}
\end{algorithmic}
\end{algorithm}

\subsection{Time Complexity}
We discuss the time complexity of the proposed method, hWMF, here. Obviously, the complexity burden of hWMF depends mostly on the computation of the derivatives in Equations~\ref{eq:partialU} and ~\ref{eq:partialV}. In each iteration in our algorithm~\ref{alg:alg1}, the time complexities of the computation of the derivatives in lines 7 and 9 are calculated as follows: first note that $\textbf{W}\odot \textbf{X}$, $\textbf{UV}^T$ and $\textbf{(W}\odot(\textbf{UV}^T))$ need to be calculated once for both equations. The time complexity of $\textbf{W}\odot \textbf{X}$ is $\mathcal{O}(N_x)$ where $N_x$ is the number of non-zero elements of the sparse matrix $\textbf{X}$. Also, the time complexity of $\textbf{UV}^T$ is $\mathcal{O}(NdM)$. For $\textbf{(W}\odot \textbf{X)V}$, we need $\mathcal{O}(N_xd)$. For the second term in Eq.~\ref{eq:partialU}, i.e. $\textbf{(W}\odot(\textbf{UV}^T))\textbf{V}$,  we need $\mathcal{O}(NM + NMd)$. Thus in each iteration, the calculation of $\frac{\partial \textbf{L}}{\partial \textbf{U}}$ takes $\mathcal{O}(N_xd + NMd)$. With the similar computations, the calculation of $\frac{\partial \textbf{L}}{\partial \textbf{V}}$ has the time complexity of $\mathcal{O}(N_xd + NMd)$ in each iteration.

In the next section, we provide statistical evidence that users are indeed more willing to adopt hashtags that have correlations with each other and consequently maintain consistent hashtag usage history. This confirms the correctness of our intuition on both incorporating the indicator matrix $\textbf{W}$ into the optimization equation and solving the equation via matrix factorization model as collaborative-filtering based models captures well the relations between items (here, hashtags).

\subsection{\textit{t}-test}
We perform two-sample \textit{t}-test and verify the existence of hashtag usage consistency. In particular, we seek to answer the question: \textit{Do users in Twitter possess consistent hashtag usage history?} 

We construct two vectors $hc_u$ and $hc_r$ with the equal number of elements where each element in $hc_u$ is obtained by calculating the correlation between hashtags $h_i$ and $h_j$ used by user $u$ using Eq.~\ref{eq:corr} and similarly each element in $hc_r$ is the calculated correlation score between hashtags $h_i$ used by user $u$ and $h_j$ used by a random user $r$. 

We perform a \textit{t}-test on vectors $hc_u$ and $hc_r$. The null hypothesis here is that the correlation of hashtags used by a given user does not differ from those of different users, i.e. these two vectors are the same, $H_0: hc_u = hc_r$, while the other hypothesis is that the hashtags used by the same user are more correlated than those used by different users, $H_1: hc_u > hc_r$. Therefore we have the following two-sample \textit{t}-test:
\begin{equation}
	H_0: hc_u = hc_r, H_1: hc_u > hc_r
\end{equation}
The \textit{t}-test result suggests a strong evidence with the significance level $\alpha = 0.001$ to reject the null hypothesis and as a result, confirms that users tend to use a consistent set of hashtags while posting tweets on Twitter. Therefore the answer to the above question is positive. This aligns well with our findings and equations in our matrix factorization model in the previous section.

\section{Experiments}
We first collect and build our dataset by making use of the Twitter streaming API\footnote{https://dev.twitter.com/streaming/overview} which provides 1\% random tweets from the total volume of tweets at a particular moment. We then conduct experiments to compare the performance of our proposed method with the baselines. In this section, we begin by introducing our dataset and the evaluation metric and then we design experiments and discuss the results.
\subsection{Dataset}
In general, the evaluation of hashtag recommendation approach is challenging due to difficulty of collecting the appropriate and standard dataset, while also human annotation is almost impossible and unreliable because of the tedious workload for evaluating data. Therefore we need a more systematic way of generating the dataset.

We rather collect the data in the following way: we first picked a set of 25 trending hashtags of different categories as our initial seeds list including: News, Obama, Iran, Yemen, Gaza, Islam, Terrorism, Shooting, BlackLivesMatters, Youtube, Apple, Google, Microsoft, iPad, Android, Internet, FIFA, JohnNash, LadyGaga, Movie, Weekend, MemorialDay, Love, Hate, Care. We then expanded this list to the 6,814 related hashtags by collecting a set of tweets containing \textit{at least} five related hashtags to the hashtags in the seeds list (see Table~\ref{tb:hashtags} for some examples). This way, we obtain a coarse-grained dataset of very close related hashtags which ease the task of evaluation of the proposed method. The description of the resulting dataset with \%0.99 spareness is shown in Table~\ref{tb:data}. 

\begin{table}
\centering
\caption{Description of the dataset}
\label{tb:data}
\begin{tabular}{l c}\hline\hline
\textbf{\# of users} & 2,976 \\
\textbf{\# of tweets} & 6,503 \\
\textbf{Max \# of paired hashtags adopted by users} & 108 \\
\textbf{Min \# of paired hashtags adopted by users} & 5 \\
\textbf{Max \# of times given paired hashtag used} & 313 \\
\textbf{Min \# of times given paired hashtag used} & 0 \\
\textbf{\# of seeds} & 25\\
\textbf{\# of hashtags} & 6,814\\\hline
\end{tabular}
\end{table}

\begin{table*}
\centering
\caption{Examples of seed hashtags and their related hashtags}
\label{tb:hashtags}
\begin{tabular}{l l}\hline
\textbf{Seed} & \textbf{Related hashtags} \\\hline
\textbf{\#Obama}& Democrats, Whitehouse, tlot, tcot, Congress \\
\textbf{\#Iran} & IranTalks, IranDeal, Nuclear, Iraq, Syria  \\
\textbf{\#iPad} & Android, iOS, iPhone, freeaps, Apple \\
\textbf{\#Google} & io15, Chrome, YouTube, Googleplay, Wearable\\\hline
\end{tabular}
\end{table*}

\begin{figure}
\centering
\includegraphics[width=0.5\textwidth]{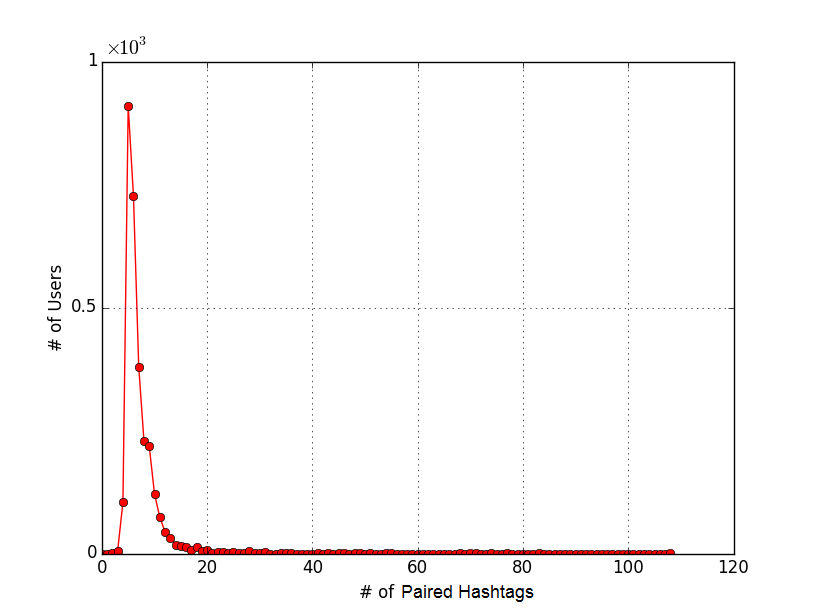}
\caption{Distribution of paired hashtags used by users}\label{fig:user_dist}
\end{figure}

\begin{figure}[t]
\centering
\includegraphics[width=0.5\textwidth]{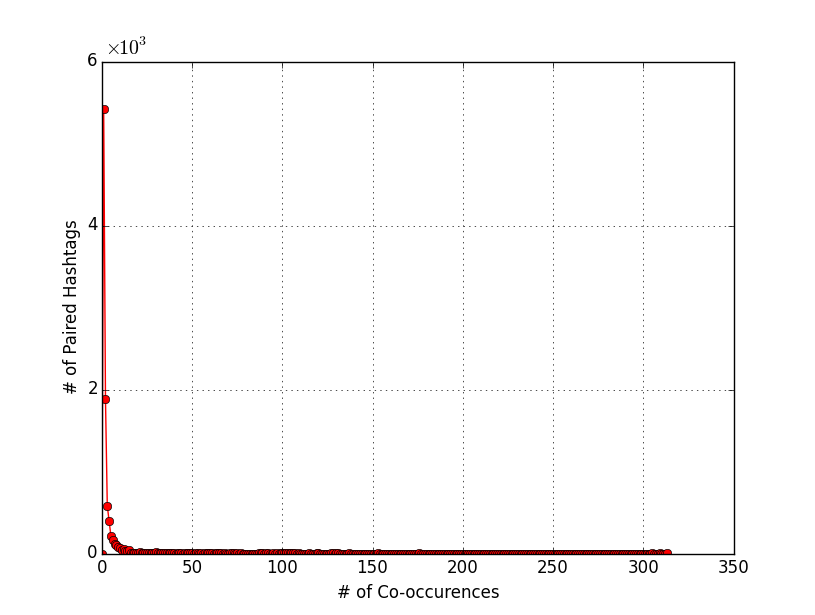}
\caption{Distribution of paired hashtags used together}\label{fig:hashtag_dist}
\end{figure}

The power-law \cite{Clauset:2009:PDE:1655787.1655789} like distributions of the paired hashtags used by users and seen together are depicted in Figures~\ref{fig:user_dist} and~\ref{fig:hashtag_dist} respectively. For the sake of clarity, we remove from Fig.~\ref{fig:hashtag_dist}, pairs of hashtags which have never appeared together. As we observe from Fig.~\ref{fig:user_dist}, very few users have used more than 20 paired hashtags while the peak in this plot shows that most of the users have only used less than 20 paired hashtags. This is somehow demonstrated in Fig.~\ref{fig:hashtag_dist} as well; roughly 5,424 paired hashtags are used together only once while we observe the decrease in the number of adopted paired hashtags as we move further in the plot. These figures together show the severe sparsity in our dataset; not all the hashtags have paired together at all; most paired hashtags are adopted together rarely. 

\begin{figure}[t]
\centering
\includegraphics[width=0.5\textwidth]{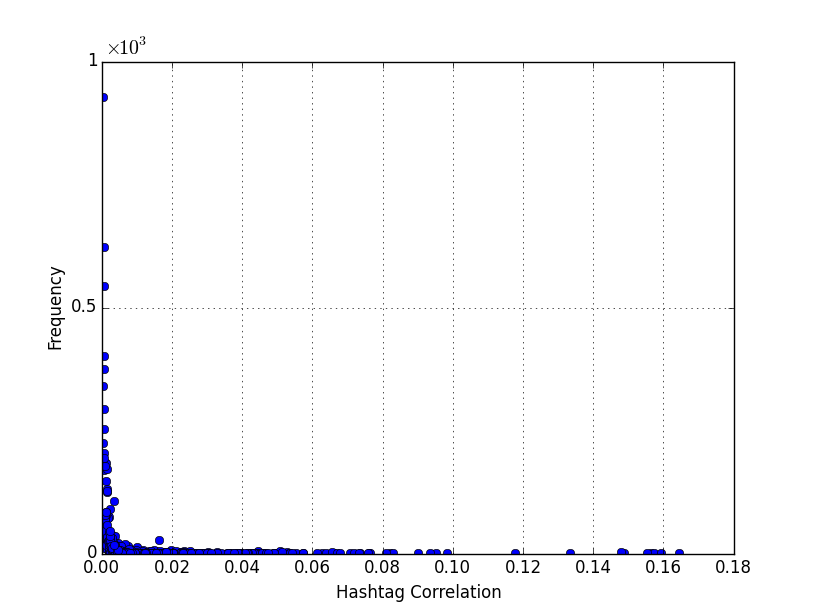}
\caption{Distribution of hashtag correlation scores.}\label{fig:corr_dist}
\end{figure}

We further plot the distribution of hashtag correlation scores in Fig.~\ref{fig:corr_dist} (based on the Eq.~\ref{eq:w}). Since the number of uncorrelated hashtags are much higher than those with correlation, once again, for clarity we remove uncorrelated hashtags. This figure demonstrates that many hashtag correlation scores fall in (0,0.1] which shows most of the hashtags do not have strong correlation with each other, suggesting a power-law like distribution of hashtag correlation scores in our dataset. 

\subsection{Evaluation Metric}
Similar to the most recommendation systems, we are unable to use popular metrics such as precision as they require to know the values of all entries in the ground truth matrix to correctly evaluate the returned values against them. Remember we had zero elements in the matrix representation of our dataset which do not necessarily show users did not adopt the hashtags, i.e. they demonstrate unknown or missing values; a user might have adopted a hashtag in the past but we could not figure it out as our data collection method does not always return all data from Twitter. Instead, we use widely used metric for evaluating collaborative filtering results, root mean square error (RMSE), which is defined as:
\begin{equation}
RMSE = \sqrt{\frac{\sum_{i,j}^n(\widetilde{x}_{ij} - x_{ij})^2}{n}}
\end{equation}

where $n$ is the number of test instances used for the evaluation and $\widetilde{x}_{ij}$ and $x_{ij}$ are corresponding test set elements of $\widetilde{X}$ and $X$ selected for the evaluation respectively.
\subsection{Discussion}
As discussed before, only few works have addressed the problem of hashtag recommendation and none of them appear to be comparable to our method since they all recommend hashtags based on the tweets contents while we do that based on the hashtag usage history. Consequently, we build two baselines and compare the results as follows,

\textbf{MF}: this is a variant of our matrix factorization model with the indicator matrix $W$ defined as follows:
\begin{equation}\label{eq:w2}
w_{ij}=\begin{cases}
1, & x_{ij} = 1\\
0, & x_{ij} = 0
\end{cases}
\end{equation}
which is basically the same as the user-hashtag matrix \textbf{X}. 

\textbf{$k$-Nearest Neighbors with correlation ($k$NN - Correlation)}: given the whole set of potential hashtags for each user, this method recommends $k$ most correlated hashtags with those each user has used before. In details, we rank hashtags according to the Eq.~\ref{eq:corr} in descending order and recommend top $k$ hashtags to the user.

\textbf{Random}: as the name suggests, this method recommends hashtags on a random basis to the users.

For our proposed method, we try various parameters and report the best performance, while other methods do not have parameters. In particular, we set both regularization parameters $\gamma_1$ and $\gamma_2$ to 0.2 and set the learning step $\lambda$ to 0.001. Also, we apply different dimensions of latent space $d$ and observe the best performance is achieved when $d=10$. For brevity, we only demonstrate this in Table~\ref{tb:results2}, when percentage of test set is fixed to \%30.

With the parameters chosen as above, our experimental setting is as follows: Suppose we have $\ell = \{ (u_i,h_j) \mid x_{ij} = 1 \}$ is the set of pairs of users and hashtags that we know they have adopted. We choose \%$x$ of $\ell$ as new relations $\mho$ between users and hashtags to predict. We remove these relations by setting $x_{ij} = 0, \forall (u_i,h_j) \in \mho$ and then apply the hashtag recommendation approaches on the new representation of \textbf{X}. We vary $x$ as $\{10,20,30,40,50\}$. 

\begin{table}
\centering
\caption{RMSE Comparison for our method with different dimensions of latent space and when percentage of test set is fixed to \%30}
\label{tb:results2}
\begin{tabular}{c|c|c|c|c}\hline
5 & 10 & 15 & 20 & 25 \\\hline\hline
0.1697  & \textbf{0.1378} & 0.1645 & 0.1709 & 0.1865 \\
\end{tabular}
\end{table}

\begin{table*}
\centering
\caption{Performance Comparison for different approaches in terms of RMSE with d = 10 and different testbed sizes}
\label{tb:results1}
\begin{tabular}{l|c|c|c|c|c}\hline
 & \textbf{\%10} & \textbf{\%20} & \textbf{\%30} & \textbf{\%40} & \textbf{\%50} \\\hline\hline
\textbf{hWMF} & \textbf{0.0913} & \textbf{0.1156} & \textbf{0.1378} & \textbf{0.1512} & \textbf{0.1693} \\
\textbf{MF} & 0.1756 & 0.1904 & 0.2023 & 0.2278 & 0.2502 \\
\textbf{$k$NN-Correlation} & 0.2827 & 0.3996 & 0.4018 & 0.4377  & 0.6229 \\
\textbf{Random} & 0.6821 & 0.6901 & 0.7045 & 0.7121 & 0.7187 \\
\end{tabular}
\end{table*}

\begin{table*}
\centering
\caption{Example hashtag usage history and suggested hashtags}
\label{tb:results3}
\begin{tabular}{l|l|l|p{8cm}}
\hline
\textbf{User} & \textbf{Hashtags adopted in past} & \textbf{Suggested hashtags} & \textbf{Tweets posted with suggested hashtags}\\\hline\hline
1 & Obama, News, Democrates & HillaryClinton & \textbf{\#HillaryClinton} how explain ALL BUSH FAULT? How still \#rap 1993-2001 \#BillClinton 100\% INNOCENT? \\
2 & Weekend, Party, Club & MemorialDay & Military Hero Joey Jones Is Honored with The Real Deal Award https://... \textbf{\#MemorialDay}
\end{tabular}
\end{table*}

We make the following observations:

\begin{itemize}
\item As stated before, for brevity, we only report the best performance of our method when the dimension of latent space $d$ is set to 10. To demonstrate the effect of $d$ on the results in Table~\ref{tb:results2}, we fix the percentage of test set to \%30 and vary $d$ as $\{5,10,15,20,25\}$. The observation suggests that $d = 10$ is the best dimension among others.

\item The performance comparison for different methods in terms of RMSE is shown in Table~\ref{tb:results1}. We observe the best results for all percentages of test set are achieved by our proposed method. The next best method is $MF$ which is a variant of the proposed framework with a different indicator matrix. One observation is that defining the indicator matrix based on the correlation between hashtags perform better than the one solely defined similar to the matrix \textbf{X}. Furthermore, the reason our method recommend hashtags more accurately than $k$NN - Correlation is because we incorporate the correlation between hashtags into widely used matrix factorizastion while $k$NN - Correlation recommends hashtags merely based on the correlation of the hashtags.

\item Table~\ref{tb:results3} shows two potential practical examples of our method for two anonymized Twitter users. As we observe, both of them have been offered one hashtag (to be concise, we do not show other recommended hashtags) and they actually adopt them in their future tweets. Obviously, these predicted hashtags are just those from the ground truth set that we use as test instances, i.e. those that we know users have previously used. This means we have not applied our method in the real world yet. 
\end{itemize}

\section{Related Work}
As opposed to the large body of works that has been devoted to the study of various social networks based applications~\cite{alvari2016identifying,beigi2014leveraging,alvari2013discovering,hajibagheri2012social,alvari2011detecting,hajibagheri2012community,beigi2016signed,beigi2016exploiting,beigi2016overview} and in particular recommendation systems in social networks~\cite{rajaraman2012mining,Bobadilla:2013:RSS:2483330.2483573,1423975, Su:2009:SCF:1592474.1722966, Massa:2007:TRS:1297231.1297235,trustAwareCollaborative2004}, very few works have addressed the problem of hashtag recommendation for easy categorization and retrieval of tweets. 

The previous works on hashtag recommendation mostly rely on the similarity between tweets and none of them have exploited the correlation between the topics of the hashtags nor the users' history of hashtags usage. In particular, they focused on two main directions: majority of them recommend hashtags based on content similarity having storage problems, fewer adopt LDA, which fails in short texts and is unsupervised and hence requires efforts to associate tweets with hashtags, and use abstracted topics of tweets. Another problem with these approaches is the contents of tweets are not always available since many users do not usually make their timelines publicly available due to the privacy concerns. Moreover, these methods suffer from computation overloads and lack of strong natural language processing techniques which are required to analyze the contents of tweets.

\cite{li2011twitter} employs WordNet similarity information and Euclidean distance to suggest hashtags from similar tweets. Given a tweet from a user, \cite{zangerle2011recommending}, first discover similar tweets in their dataset based on the TF-IDF\cite{salton1986introduction} representation of tweets. Then they recommend hashtags by three different approaches, relying on ranking the hashtags based on the overall popularity of tweets, the popularity within the most similar tweets, and third one, the most similar tweets, reported to perform the best among the others. As opposed to their method which is solely based on the tweets similarities, \cite{conf/ideas/OtsukaWC14} considers terms in tweets and their relevance to candidate hashtags. In more details, they propose a ranking scheme, a variant of TF-IDF that takes into account the relevance of hashtags and data sparsity. Similar to their method, \cite{conf/www/GodinSNSW13}, proposes an approach focusing on detection of hidden topics for the tweets and recommending the use of those general topics as hashtags based on LDA model.  In details, an unsupervised and content based hashtag recommendation based on LDA is proposed in this study. They first design and develop a binary language classifier for tweets based on the Naive Bayes and Expectation-Maximization and then apply LDA in the context of tweet hashtag recommendation in a fully unsupervised manner.

Different from the other studies, \cite{conf/www/She014} proposes a supervised topic modeling based approach for recommendation of hashtags by treating hashtags as labels of topics and leveraging topic model to discover relationship between words, hashtags and topics of tweets. with the assumption that each tweet is about one local topic and there is a global background topic for the corpus. They finally recommend $k$ most probable hashtags based on the probability that a new tweet would contain a hashtag. \cite{Mazzia_suggestinghashtags}, uses probability distributions for hashtag recommendation. In particular, they uses Bayes rule to compute the maximum a posteriori probability of each hashtag class given the words of the tweet. 

In an different attempt,~\cite{kywe2012recommending} proposes a method that recommends hashtags based on similar users and tweets by calculating the preference weight of a user towards a certain hashtag based on the TF-IDF and then selecting the top users with high cosine similarities with other users. Also, they use the same approach for calculating the top similar tweets. Since many tweets do not contain hashtags, the recommended hashtags may be from similar tweets instead of similar users. They claim that their method is able to recommend more personalized hashtags compared to the other methods and their method suits both user preferences and the tweet content.

In contrast, this study seeks to recommend hashtags by incorporating hashtag usage history of each user into the hashtag recommendation problem and address the problem with an optimization problem while using matrix factorization model to solve it. In particular, we do not rely on the network structure or similarity between tweets/users; instead we exploit the correlation between hashtags by focusing on the probability of their adoption together.

\section{Conclusion and Future Work}
In this study, we presented an approach to recommend hashtags to the users solely based on their interests and hashtags usage history, which in contrast to the approaches in the literature, does not require the contents of tweets as they are not always available since many users do not usually make their timelines publicly available due to the privacy concerns. In addition to that, all other methods suffer from severe computation burden and lack of strong natural language processing techniques to analyze the contents of tweets. We formulated the problem into an optimization problem and integrated correlation between different hashtags into the equation and used widely used matrix factorization technique to solve the problem via alternating least squares scheme. Experiments demonstrate that our method outperforms other baselines in recommending more accurate hashtags to the users.

In future, we plan to replicate the study by investigating the correlation between hashtags based on first discovering their topics from the content of their corresponding tweets through using topic modeling techniques such as LDA, and then computing their topic similarity based on the existing popular similarity measures. That way, we can integrate more accurate correlation values between hashtags into our matrix factorization model. 

We are investigating other interesting research directions such as time series analysis of the adoption of hashtags by different users and extend the model to work well in presence of time slots in order to design seasonal hashtag recommendation systems as well as incorporating demographics of users specially gender of the users to recommend personalized hashtags to them.

%
\bibliographystyle{abbrv}
\bibliography{sigproc}  
%
%
\end{document}